\documentclass[runningheads]{llncs}
\usepackage{graphicx}
\usepackage{cite}
\begin{document}
\title{Multi-scale Regional Attention Deeplab3+: Multiple Myeloma Plasma Cells Segmentation in Microscopic Images}
\author{Afshin Bozorgpour\inst{1,2\ast} \and
Reza Azad\inst{1,2\ast} \and
Eman Showkatian\inst{2} \and
Alaa Sulaiman\inst{2\dag}}

\authorrunning{A. Bozorgpour et al.}
\titlerunning{Multi-scale Regional Attention Deeplab3+}
\institute{Sharif University of Technology, Tehran, Iran \and
BmDeep Group, contact@bmdeep.com}
% \\ \textit{The two first authors are equally contributed}

%
\maketitle              % typeset the header of the contribution
\begin{abstract}
Multiple myeloma cancer is a type of blood cancer that happens when the growth of abnormal plasma cells becomes out of control in the bone marrow. There are various ways to diagnose multiple myeloma in bone marrow such as complete blood count test (CBC) or counting myeloma plasma cell in aspirate slide images using manual visualization or through image processing technique. In this work, an automatic deep learning method for the detection and segmentation of multiple myeloma plasma cell have been explored. To this end, a two-stage deep learning method is designed. In the first stage, the nucleus detection network is utilized to extract each instance of a cell of interest. The extracted instance is then fed to the multi-scale function to generate a multi-scale representation. The objective of the multi-scale function is to capture the shape variation and reduce the effect of object scale on the cytoplasm segmentation network. The generated scales are then fed into a pyramid of cytoplasm networks to learn the segmentation map in various scales. On top of the cytoplasm segmentation network, we included a scale aggregation function to refine and generate a final prediction. The proposed approach has been evaluated on the SegPC2021 grand-challenge and ranked second on the final test phase among all teams. 

\keywords{Myeloma Plasma Cell \and Segmentation \and Attention Deeplabv3+ \and Deep Learning \and SegPC2021 \and Grand Challenge.}
\end{abstract}

% \Yinyang
\let\thefootnote\relax\footnotetext{\scriptsize
\noindent
    \quad \textbf{Support:} This work is fully supported by BmDeep Group.\\
    $\dag$ \textbf{To whom correspondence should be addressed:} E-mail: alaa@bmdeep.com.\\
    $\ast$ These authors contributed equally to this work.
}

\section{Introduction}
Cancer happens when the cells start to grow out of control and spread to healthy surrounding tissue. Myeloma, also known as multiple myeloma, is a type of blood cancer that arises from plasma cells in the bone marrow \cite{rajkumar2014international, guyton2006medical}. More specifically, bone marrow is a kind of soft tissue found inside some part of larger bones in human body. Different types of blood cells such as red blood cells, white blood cells, and platelets are made in the bone marrow \cite{hideshima2007understanding}. Plasma cells developed by the B lymphocytes (type of white blood cells) form part of the body's immune system. In order to fight the infections, antibodies, also known as immunoglobulin, are produced by normal plasma cells. In myeloma cancer plasma cells crow in the bone marrow in a way there is no space for normal red cells, white cells, and platelets. Myeloma begins to develop when the DNA is damaged or changed during the production of new plasma cells. These abnormal plasma cells (myeloma cells) will spread in a different part of bone marrow and produce more abnormal cells. Myeloma cells will produce a large number of paraproteins (type of antibody) which are useless and unable to fight the infections \cite{bird2011guidelines}. Unlike other cancers, myeloma will not form a tumor or lump but it will lead to the accumulation of abnormal plasma cells in the bone marrow and paraproteins in the body. Multiple myeloma is referred to the situation when myeloma cancers affect multiple parts of the body \cite{alexanian1994treatment}. Since myeloma cell can be differentiated from normal plasma cells based on histology and morphological features, it is a common method to diagnose multiple myeloma cancer through the aspirate slide images \cite{kyle2007clinical, palumbo2009international, nau2008multiple}. In this method, at first, blood samples will be extracted from bone marrow by using the injection of the needle onto the bone. The extracted blood sample will be transferred to a slide and stained using hematoxylin and eosin. Abnormal plasma cells will be detected and marked using manual microscopic visualization (sample is shown in figure 1). Finally, based on the estimation of the normal plasma cells in bone marrow, the presence or absence of myeloma cancer will be concluded \cite{minges2001plasma}.

\begin{figure}
\includegraphics[width=\textwidth]{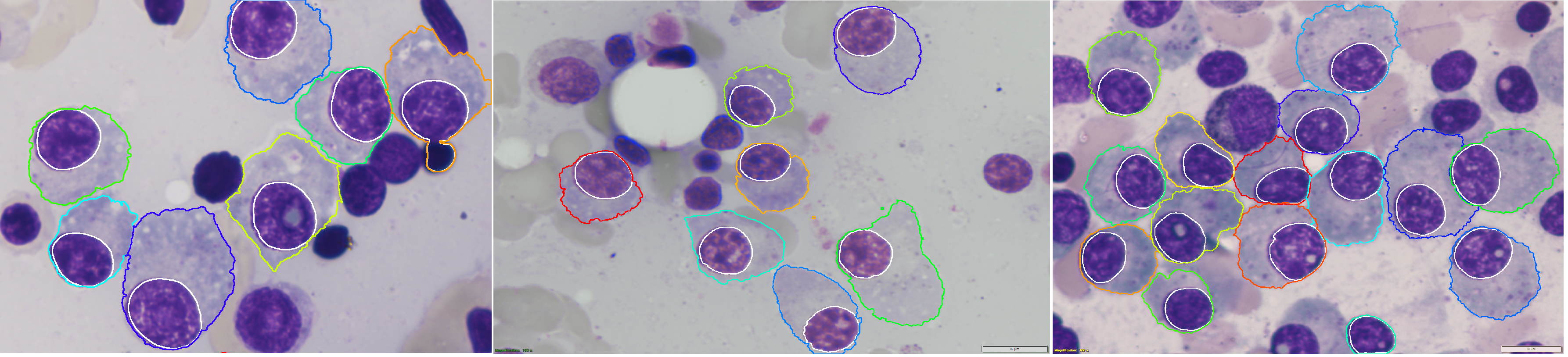}
\caption{Some samples of myeloma plasma cells microscopy images \cite{segpc2021}, where the myloma cancer cells are detected and highlighted with colored boundary.} \label{fig:figure1}
\end{figure}

Although manual inspection of stained slide images is a gold standard of diagnosis of myeloma cancer, it is time-consuming and prone to inter and intraobserver variation. These limitations could be compensated by the use of advanced digital image processing techniques such as object detection and segmentation. Automation of the abnormal plasma cell detection alongside expert pathologist decision could lead to the reduction of diagnosis time and workload of the pathologist. To this end, in this paper, we propose a deep network that utilizes a pyramid of Attention Deeplabv3+ model in a regional-based manner to segment each instance of a myeloma cancer cell. We utilize our approach on the multiple mylomia plasma cell segmentation challenge which provided by \cite{gupta2018pcseg,gupta2020gcti, gehlot2020ednfc, segpc2021}. Our contribution is summarized as follows:
\begin{itemize}
    \item Second ranking on the SegPC2021 challenge for multiple myeloma cancer cell segmentation.
    \item Regional base instance segmentation approach.
    \item guiding cytoplasm segmentation network with additional nucleus mask as a supervisory signal.
\end{itemize}

\section{Related Work}
Advanced image processing and machine learning methods such as image classification, object detection, and segmentation could have promising applications in various medical domain \cite{azad2019bi, asadi2020multi, feyjie2020semi}. Detection and segmentation of abnormal cells in microscopic images have been proposed by several researchers in recent works. For instance, Vyshnav et al. used a deep learning-based approach for multiple myeloma cancer detection in stained microscopic images. They compared the performance of Mask R-CNN and U-Net in segmentation and inferred that Mask R-CNN has a very better performance than U-Net in myeloma cell segmentation \cite{vyshnav2020deep}. Authors of \cite{tehsin2019myeloma} used convolutional neural networks to classify the normal and abnormal plasma cells in stained microscopic images. For the classification task, they used AlexNet \cite{krizhevsky2012imagenet} to extract features from microscopic images and then used Support Vector Machine (SVM) for the classifier.  Their novelty was the preprocessing stage where they used a median filter for each R, G, and B color channel (on the importance of color space \cite{azad2014novel}) individually and linear contrast stretching for the color enhancement. Then they compared their result for classification with state of art techniques. They reported that their networks reached $89\%$ sensitivity for the cell classification. Vuola et al. used Mask-RCNN and U-Net ensembled for nuclei segmentation in microscopic images. They inferred that Mask-RCNN and U-Net have similar results on the nucleus segmentation task. They reported that U-Net has better performance in nucleus segmentation than Mask R-CNN in the term of similarity index. On the other hand, the Mask R-CNN has better performance in the term of precision assessment. Finally, they concluded that an ensembled model improves the model performance in nucleus segmentation \cite{vuola2019mask}. Saeedizadeh et al. used a bottleneck algorithm, modified watershed, and SVM for myeloma cell detection in microscopic images. At first, they separated the white blood cell from red blood cells by use of the color normalization. After that the used thresholding technique to separate the nucleus from the cytoplasm and by the use of watershed and bottleneck algorithm they separated the connected cells. Finally, by using the series of decision rules and the use of an SVM classifier they achieved the sensitivity of $96.52\%$ and precision of $95.28\%$ in recognition of myeloma cells \cite{saeedizadeh2016automatic}. Even though the literature work gained promising results, their applicability in the multiple myeloma cancer cell segmentation is limited due to the challenges such as overlap between cytoplasm's of instances, the fuzzy boundary of the cytoplasm, and overlaying of one nucleus on another cytoplasm in microscopic images. To mitigate these limitations, we propose to regional attention deep model to segment each cell with precise attention.

\section{Methodology}
A general diagram of the proposed structure is depicted in figure 2. The proposed methods consist of two stages: in the first stage, the nucleus segmentation network extracts all the nucleus instances. Then each instance fed into a multiscale cytoplasm segmentation network. This network utilizes the Attention Deeplab3+ model to segment the cytoplasm area. In the next subsections, we will elaborate on each part in more detail.

\begin{figure}
\includegraphics[width=\textwidth]{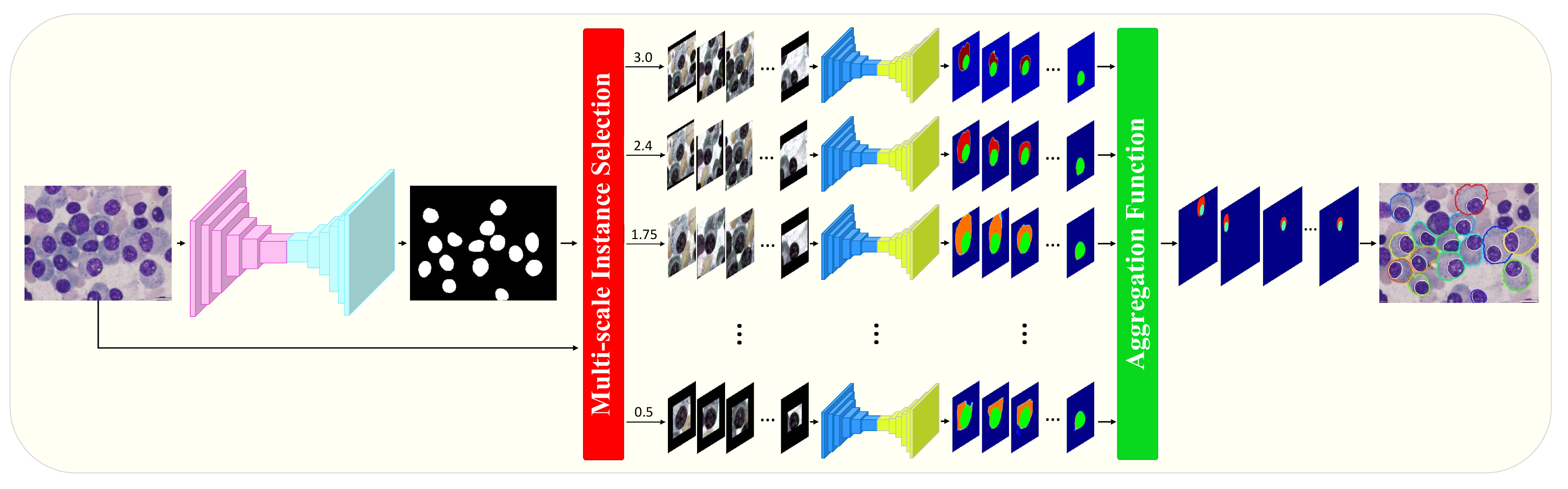}
\caption{Proposed regional Attention Deeplabv3+ model for multiple myeloma plasma cells segmentation. The proposed method applies a U-net structure to learn the segmentation map for each nucleus instance then it utilizes multi-scale attention deeplabv3+ model to generate the segmentation mask for cytoplasm.} \label{fig:figure1}
\end{figure}

\subsection{Nucleus Segmentation}
In the proposed architecture instead of jointly learning the segmentation of the nucleus and cytoplasm mask, we utilize a two-stage strategy. Our main motivation is to use the detected nucleus instance as a supervisory signal for the cytoplasm instance segmentation to deal with overlapped areas. In other words, the man objective of the first stage is to extract all the possible nucleus instances from the input image. Then each extracted nucleus instance alongside the cropped image patch is fed to the multi-scale instance selection function. The instance selection function is simply an image cropping function with a predefined scale. We use multi-scale to deal with varying cytoplasm scales.  In figure 2, a sample of cropped nucleus instances with varying scale sizes ($0.5$ to $3.0$) is demonstrated. To learn the nucleus segmentation map we train a U-net model using a nucleus annotation mask. It is worthwhile to mention that we include the predicted nucleus instance alongside the cropped image as an input for the cytoplasm segmentation network. The goal of this extra input is to guide the network for the object of interest.  

\subsection{Cytoplasm Segmentation}
\begin{figure}
\rotatebox{0}{\scriptsize \hspace{0.8cm} Input \hspace{0.85cm} GT Mask \hspace{.5cm} Estimation \hspace{0.88cm} Input \hspace{0.85cm} GT Mask \hspace{.5cm} Estimation}
\\\rotatebox{90}{\scriptsize \hspace{.25cm} Scale 2.6 \hspace{.8cm} Scale 1.0}
\includegraphics[width=0.97\textwidth]{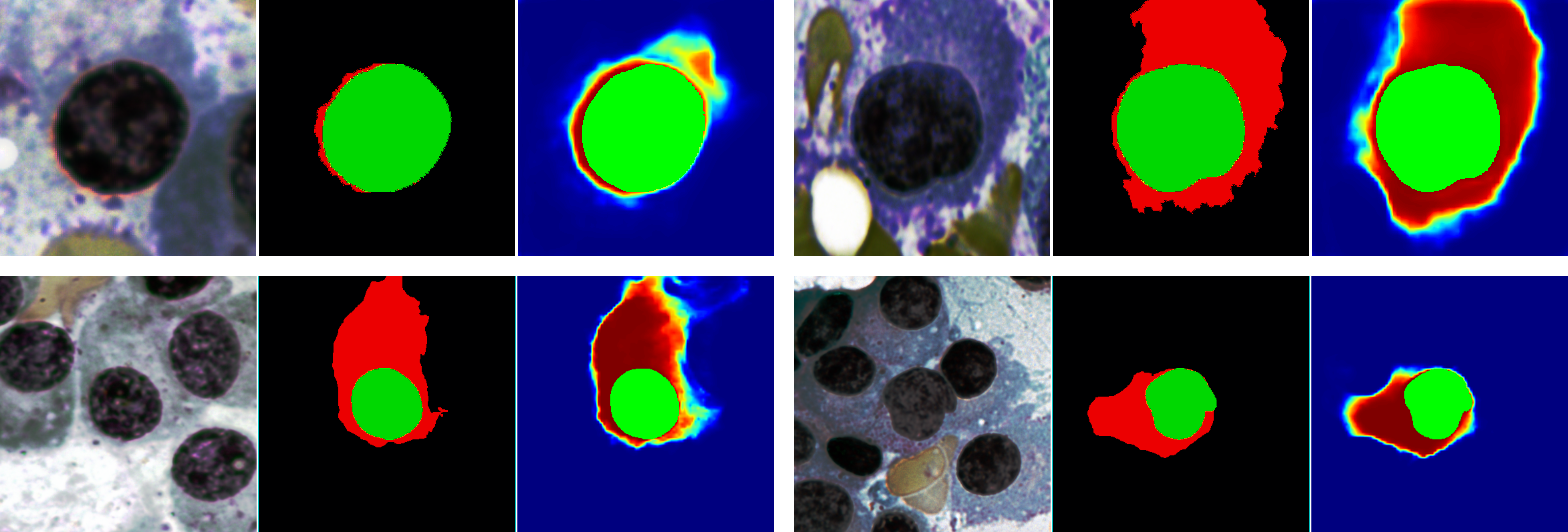}
\caption{Segmentation results of the proposed method for both nucleus and cytoplasm area. The cropped image alongside the predicted nucleus mask is fed to the cytoplasm network to generate the instance cytoplasm segmentation.} \label{fig:experimental_inferences}
\end{figure}

In a regular auto-encoder decoder structure the encoder network consists of several convolutions blocks followed by pooling operations to encode the object of interest in high-level representation space.
In this structure, due to the consecutive pooling operation, the spatial dimension of the network may considerably decrease which can result in less discriminated representation power for objects with varying scale. To mitigate this problem, the Deeplabv3+ model utilized an atrous convolution structure. The atrous convolution applies a set of upsampled convolutional kernels to describe the object of interest in higher receptive filed size. To further improve the representation power of the Deeplab model, Azad et al. \cite{azad2020attention} proposed a two-level add-on attention mechanism to extract more informative features from the atrous convolutions. Where the first level attention mechanism scales the representation space to highlight the more informative channels then the second attention mechanism utilizes a 3D convolution kernel between each atrous scale to learn a robust non-learn feature set.
In this section, we use the Attention Deeplabv3+ model to tackle the cytoplasm segmentation problem. Cytoplasm boundary has a high overlap with the background area and it requires careful attention to discriminate the cytoplasm boundary from the background area. In our implementation, we fed the extracted nucleus area alongside the image batch to the model to learn the instance segmentation mask. We also apply the image histogram equalization method to normalize samples. Sample of the estimated masks for the given nucleus instance is depicted in figure \ref{fig:experimental_inferences}.

\subsection{Aggregation function}
Learning objects of interest in multi-scale fashion can produce a robust segmentation mask. In this work, we apply the multi-scale technique on the input level. Consequently, the model generates a multi-scale segmentation mask. The main objective of this multi-scale technique is to tackle the problem of cytoplasm boundary. More specifically, the cytoplasm boundary has a non-rigid shape. If the boundary area is less then the model requires precise attention around the nucleus to separate it from the background. On the other hand, if the boundary is wast then the model needs to consider the big area to separate its boundary from other instances. We solve this limitation by defining several scales.  
Since the ultimate objective of the model is to produce a single segmentation mask for each instance, we propose to use an aggregation function to combine and select a single segmentation mask. The aggregation function can use the output of all scales to generate a single prediction (like non-maxima suppression), however, our experiment selection approach produced better performance. To perform this operation, we simply start from the lowest scale and calculate the relation between the detected cytoplasm area and the nucleus area. If the ratio is higher than a threshold value then the next scale is evaluated. The process goes through the next scales until finding the appropriate condition.

\subsection{Training Procedure}
Our training procedure consists of two stages. In the first stage, we train the U-net model to segment the nucleus from the input images. The training process takes into account the training and validation set and learns the nucleus mask. We train the model for 100 epochs using the Adam optimization with a learning rate of $1e-4$. In the second phase, we trained each Attention Deeplabv3+ model using the patches extracted from the input image alongside the nucleus mask (resulted from the nucleus segmentation network). For each scale, we train the model for 100 epochs using the Adam optimization with a learning rate of $1e-5$. All training is done using cross-entropy loss on a single GTX 1080 GPU.

\subsection{Inference Procedure}
The inference stage uses the trained models to generate the segmentation mask for both nucleus and cytoplasm instances. In our inference, we use a fixed number of scales (4 scales) for the test phase. 

\section{Results}
The proposed method is evaluated on multiple myeloma cell segmentation grand challenges which are provided by the SegPC 2021. The challenge data set consists of a training set with 290 samples, validation and test sets with 200 and 277 samples respectively. All the samples are annotated by the pathologist and instance base segmentation masks are provided for the object of interest (myeloma plasma cells). We trained our model using the training and validation set. During the competition time, we generated the segmentation mask for each instance. The challenge leader-board compared each team using the MIOU metric, where our method ranked second among all teams. Table 1 shows the comparison results for the top five winning teams. 

\begin{table}[!ht]
\centering
    \caption{Performance comparison on the final test phase for SegPC2021 challenge.}
% 	\resizebox{0.3\columnwidth}{!}{
	\begin{tabular}{c|c}
		\hline
		\textbf{Teams} & \textbf{Score (mIoU)}\\
		\hline
		XLAB Insights  & 0.9389 \\
		\textbf{bmdeep}  & \textbf{0.9385} \\
		DSC-IITISM  & 0.9382 \\
		507  & 0.9366 \\
		AIVIS  & 0.9276 \\
		\hline
	\end{tabular}
% 	}
	\label{tab:ftpl}
\end{table}

As shown in table 1, the proposed method outperformed most of the competitors and achieved the second-best place with a small gap ($0.0004$) from the first team. Figure 4 demonstrates some prediction results where the proposed method estimated both nucleus and cytoplasm masks with high performance.

\begin{figure}
{\rotatebox{0}{\scriptsize \hspace{1.7cm} Input \hspace{2.6cm} Ground-Truth \hspace{2.35cm} Estimation}}\\
\includegraphics[width=\textwidth]{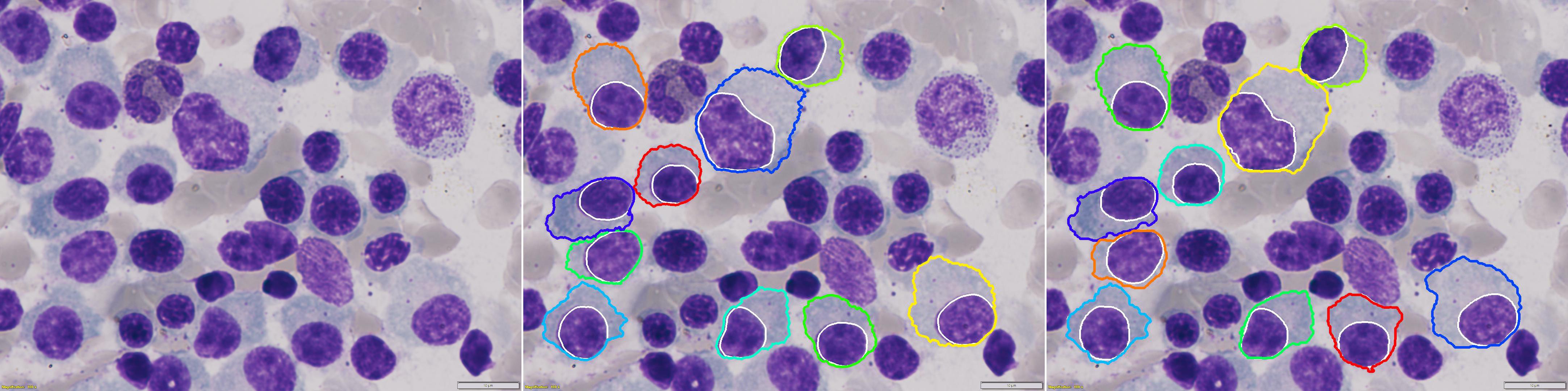}
\caption{A sample of prediction results on SegPC2021 grand challenge.} \label{fig:fig4}
\end{figure}

As explained before the proposed method uses a multi-scale strategy to generate a precise segmentation mask for the cytoplasm instances. 
In this section, we will elaborate on the scale selection strategy and its effect on the final performance. To this end, we have extracted the statistical scale information from the training set. The histogram information is depicted in figure 5. 

\begin{figure}[!ht]
\centering
\includegraphics[width=0.5\textwidth]{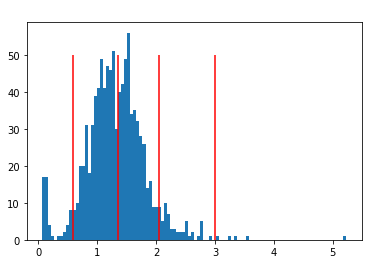}
\caption{Histogram of the area of cytoplasm to nucleus on the validation set.} \label{fig:fig5}
\end{figure}

According to figure \ref{fig:fig5}, it is crystal clear that the ratio of cytoplasm area to nucleus area is distributed in almost four different peaks. Thus, we select four scales to generate a precise segmentation mask. It is worthwhile to mention that we select the scale value a bit higher than the histogram peaks (shown with red line on figure 5) to generate an image patch to cover the appropriate receptive field size. In our experiment for the final test phase, we selected 4 different scales as depicted in figure 2.

\section{Conclusion}
In this paper, we proposed a multi-scale regional Attention Deeplabv3+ model for myeloma plasma cell segmentation. The proposed method utilized a U-net model for nucleus instance segmentation. The segmented nucleus instance is extracted from the input image and alongside the predicted nucleus mask fed into a multi-scale cytoplasm detector network. The cytoplasm detector took into account the strength of the Attention Deeplabv3+ model to segment each cytoplasm instance. We further proposed an aggregation function to select the more related scale to fulfill the prediction score. Evaluation results on the final challenge phase demonstrated outstanding results. 

\hspace{0.5cm} \\ \noindent
\textbf{Acknowledgements}. This work was fully supported by the BmDeep group. All the implementation code is available: https://github.com/bmdeep/SegPC2021

\bibliographystyle{splncs04}
\bibliography{ref}

\begin{thebibliography}{10}
\providecommand{\url}[1]{\texttt{#1}}
\providecommand{\urlprefix}{URL }
\providecommand{\doi}[1]{https://doi.org/#1}

\bibitem{alexanian1994treatment}
Alexanian, R., Dimopoulos, M.: The treatment of multiple myeloma. New England
  Journal of Medicine  \textbf{330}(7),  484--489 (1994)

\bibitem{asadi2020multi}
Asadi-Aghbolaghi, M., Azad, R., Fathy, M., Escalera, S.: Multi-level context
  gating of embedded collective knowledge for medical image segmentation. arXiv
  preprint arXiv:2003.05056  (2020)

\bibitem{azad2019bi}
Azad, R., Asadi-Aghbolaghi, M., Fathy, M., Escalera, S.: Bi-directional
  convlstm u-net with densley connected convolutions. In: Proceedings of the
  IEEE/CVF International Conference on Computer Vision Workshops. pp.~0--0
  (2019)

\bibitem{azad2020attention}
Azad, R., Asadi-Aghbolaghi, M., Fathy, M., Escalera, S.: Attention deeplabv3+:
  Multi-level context attention mechanism for skin lesion segmentation. In:
  European Conference on Computer Vision. pp. 251--266. Springer (2020)

\bibitem{azad2014novel}
Azad, R., Shayegh, H.R.: Novel and tuneable method for skin detection based on
  hybrid color space and color statistical features. arXiv preprint
  arXiv:1407.6506  (2014)

\bibitem{bird2011guidelines}
Bird, J.M., Owen, R.G., D’Sa, S., Snowden, J.A., Pratt, G., Ashcroft, J.,
  Yong, K., Cook, G., Feyler, S., Davies, F., et~al.: Guidelines for the
  diagnosis and management of multiple myeloma 2011. British journal of
  haematology  \textbf{154}(1),  32--75 (2011)

\bibitem{feyjie2020semi}
Feyjie, A.R., Azad, R., Pedersoli, M., Kauffman, C., Ayed, I.B., Dolz, J.:
  Semi-supervised few-shot learning for medical image segmentation. arXiv
  preprint arXiv:2003.08462  (2020)

\bibitem{gehlot2020ednfc}
Gehlot, S., Gupta, A., Gupta, R.: Ednfc-net: Convolutional neural network with
  nested feature concatenation for nuclei-instance segmentation. In: ICASSP
  2020-2020 IEEE International Conference on Acoustics, Speech and Signal
  Processing (ICASSP). pp. 1389--1393. IEEE (2020)

\bibitem{gupta2020gcti}
Gupta, A., Duggal, R., Gehlot, S., Gupta, R., Mangal, A., Kumar, L., Thakkar,
  N., Satpathy, D.: Gcti-sn: Geometry-inspired chemical and tissue invariant
  stain normalization of microscopic medical images. Medical Image Analysis
  \textbf{65},  101788 (2020)

\bibitem{segpc2021}
Gupta, A., Gupta, R., Gehlot, S., Goswami, S.: Segpc-2021: Segmentation of
  multiple myeloma plasma cells in microscopic images. IEEE Dataport
  \textbf{1}(1), ~1 (2021)

\bibitem{gupta2018pcseg}
Gupta, A., Mallick, P., Sharma, O., Gupta, R., Duggal, R.: Pcseg: Color model
  driven probabilistic multiphase level set based tool for plasma cell
  segmentation in multiple myeloma. PloS one  \textbf{13}(12),  e0207908 (2018)

\bibitem{guyton2006medical}
Guyton, A.C., Hall, J.E.: of medical (2006)

\bibitem{hideshima2007understanding}
Hideshima, T., Mitsiades, C., Tonon, G., Richardson, P.G., Anderson, K.C.:
  Understanding multiple myeloma pathogenesis in the bone marrow to identify
  new therapeutic targets. Nature Reviews Cancer  \textbf{7}(8),  585--598
  (2007)

\bibitem{krizhevsky2012imagenet}
Krizhevsky, A., Sutskever, I., Hinton, G.E.: Imagenet classification with deep
  convolutional neural networks. Advances in neural information processing
  systems  \textbf{25},  1097--1105 (2012)

\bibitem{kyle2007clinical}
Kyle, R.A., Remstein, E.D., Therneau, T.M., Dispenzieri, A., Kurtin, P.J.,
  Hodnefield, J.M., Larson, D.R., Plevak, M.F., Jelinek, D.F., Fonseca, R.,
  et~al.: Clinical course and prognosis of smoldering (asymptomatic) multiple
  myeloma. New England Journal of Medicine  \textbf{356}(25),  2582--2590
  (2007)

\bibitem{minges2001plasma}
Minges~Wols, H.A.: Plasma cells. e LS  (2001)

\bibitem{nau2008multiple}
Nau, K.C., Lewis, W.D.: Multiple myeloma: diagnosis and treatment. American
  family physician  \textbf{78}(7),  853--859 (2008)

\bibitem{palumbo2009international}
Palumbo, A., Sezer, O., Kyle, R., Miguel, J., Orlowski, R., Moreau, P.,
  Niesvizky, R., Morgan, G., Comenzo, R., Sonneveld, P., et~al.: International
  myeloma working group guidelines for the management of multiple myeloma
  patients ineligible for standard high-dose chemotherapy with autologous stem
  cell transplantation. Leukemia  \textbf{23}(10),  1716--1730 (2009)

\bibitem{rajkumar2014international}
Rajkumar, S.V., Dimopoulos, M.A., Palumbo, A., Blade, J., Merlini, G., Mateos,
  M.V., Kumar, S., Hillengass, J., Kastritis, E., Richardson, P., et~al.:
  International myeloma working group updated criteria for the diagnosis of
  multiple myeloma. The lancet oncology  \textbf{15}(12),  e538--e548 (2014)

\bibitem{saeedizadeh2016automatic}
Saeedizadeh, Z., Mehri~Dehnavi, A., Talebi, A., Rabbani, H., Sarrafzadeh, O.,
  Vard, A.: Automatic recognition of myeloma cells in microscopic images using
  bottleneck algorithm, modified watershed and svm classifier. Journal of
  microscopy  \textbf{261}(1),  46--56 (2016)

\bibitem{tehsin2019myeloma}
Tehsin, S., Zameer, S., Saif, S.: Myeloma cell detection in bone marrow
  aspiration using microscopic images. In: 2019 11th International Conference
  on Knowledge and Smart Technology (KST). pp. 57--61. IEEE (2019)

\bibitem{vuola2019mask}
Vuola, A.O., Akram, S.U., Kannala, J.: Mask-rcnn and u-net ensembled for nuclei
  segmentation. In: 2019 IEEE 16th International Symposium on Biomedical
  Imaging (ISBI 2019). pp. 208--212. IEEE (2019)

\bibitem{vyshnav2020deep}
Vyshnav, M., Sowmya, V., Gopalakrishnan, E., VV, S.V., Menon, V.K., Soman, K.:
  Deep learning based approach for multiple myeloma detection. In: 2020 11th
  International Conference on Computing, Communication and Networking
  Technologies (ICCCNT). pp.~1--7. IEEE (2020)

\end{thebibliography}

\end{document}